\documentstyle[proceedings,epsf]{crckapb}

\begin{opening}
\title{Fragmentation of Molecular Clouds: The Initial Phase of a
Stellar Cluster}
\author{Ralf S. Klessen}
\author{Andreas Burkert}
\institute{Max-Planck-Institut f{\"u}r Astronomie\\
K{\"o}nigstuhl 17, 69117 Heidelberg, Germany\\
klessen@mpia-hd.mpg.de \\ burkert@mpia-hd.mpg.de
}
\end{opening}

\runningtitle{Fragmentation in Molecular Clouds}
 
\begin{document}

\begin{abstract}
Using smoothed particle hydrodynamics in combination with the
special-purpose hardware device {\sc Grape}, we follow the
fragmentation process and dynamical evolution in the interior of
molecular clouds until most of the gas is converted to form a 
cluster of protostellar cores. The properties of this protostellar
cluster are compared to observational data from star forming regions.
\end{abstract}

\section{Introduction}
The dynamical evolution and fragmentation in molecular clouds is a
very complex process leading to the formation of stars. It involves a
wide range of physical phenomena and spans many orders of magnitude in
gas density. Molecular clouds have very complicated structure.
Observations reveal a hierarchy of clumps and sub-clumps on all
scales accessible by today's telescopes.

To examine this problem, we start in the most basic way: We follow the
evolution of a region in the interior of a molecular
cloud using smoothed particle hydrodynamics (Benz 1990, Monaghan 1992)
in combination with the special-purpose hardware device {\sc Grape}
(Sugimoto et al. 1990, Ebisuzaki et al. 1993). Periodic boundary conditions prevent
global collapse of the system (Klessen 1997). {\sc Grape} solves Poisson's
equation by direct summation and returns a list of nearest neighbors
for each particle, which makes it suitable for combination with SPH.
To start with, we adopt a simple isothermal
model, and thus concentrate on the interplay between gravity and gas pressure.
This produces a hierarchical network of filaments, sheets and collapsing
protostellar knots. In our simulations, once a collapsed core has formed,
we replace it by a `sink' particle that has the ability to
accrete from its surrounding gaseous envelope (Bate et al. 1995). Starting from Gaussian
initial density perturbations we follow the evolution of the system until most
gas is consumed by the newly formed cluster of protostars.

Our goal is to find a description of how the formation of these objects and
their properties depend on the initial conditions. Hence,
we follow the fragmentation process for a range of {\em temperatures}
and different sets of {\em initial density
fields} (typically Gaussian with varying power laws or turbulent velocity
structure).

These clusters of accreting protostellar clumps have to be compared with
the observational data in star forming regions: Mass spectra, kinematical
properties, multiplicity and spatial distribution can be used to constrain
the physical model  and exclude regions in parameter space that cannot
reproduce the observations.

\begin{figure}[ht]
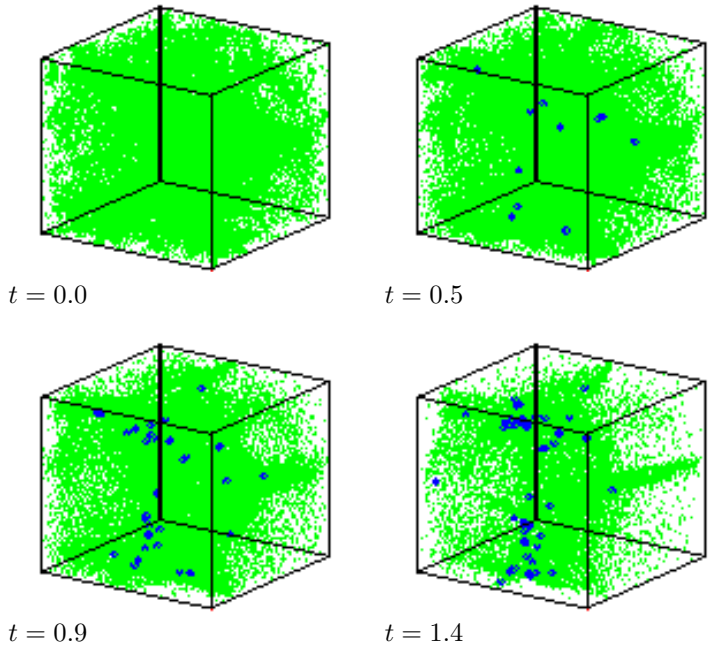

\label{fig:a}
\unitlength1cm
\begin{picture}(10,9)
\put( 1.6, 4.6){\small $t=0.0$}
\put( 6.6, 4.6){\small $t=0.5$}
\put( 1.6, 0.1){\small $t=0.9$}
\put( 6.6, 0.1){\small $t=1.4$}
\put( 1.2, 4.1){\epsfxsize=5.5cm \epsfbox{klessen-fig-1a}}
\put( 6.2, 4.1){\epsfxsize=5.5cm \epsfbox{klessen-fig-1b}}
\put( 1.2,-0.4){\epsfxsize=5.5cm \epsfbox{klessen-fig-1c}}
\put( 6.2,-0.4){\epsfxsize=5.5cm \epsfbox{klessen-fig-1d}}
\put( 1.6, 4.6){\small $t=0.0$}
\put( 6.6, 4.6){\small $t=0.5$}
\put( 1.6, 0.1){\small $t=0.9$}
\put( 6.6, 0.1){\small $t=1.4$}
\end{picture}
\caption{ Time evolution and fragmentation of a region of 222 Jeans
masses in the interior of a molecular cloud with initial Gaussian
density fluctuations having a power law of $P(k) \propto 1/k^2$.
Collapse sets in and soon forms a cluster of highly condensed cores,
which continue to accrete from the surrounding gas reservoir. At
$t=0.5$ (measured in terms of the free-fall time of the isolated gas
cube) about 10\% of all the gas mass is converted into protostellar
cores, shown as dark points. At $t=0.9$ about 30\% of all gas is
accreted and at $t=1.4$ this value is 65\%. }
\end{figure}

\section{A Case Study}
As an example, we present in Fig.~\ref{fig:a} the time evolution and
fragmentation of a region of 222 Jeans masses in the interior of a
molecular cloud with initial Gaussian density fluctuations of power
law $P(k) \propto 1/k^2$.  Pressure smears out small scale features,
whereas large scale fluctuations start to collapse into filaments and
knots. After about one quarter of the overall free-fall time, the
first highly collapsed cores appear and more follow. A hierarchically
structured cluster of accreting protostellar objects forms.

Isothermal gas cannot heat, accretion continues until the entire gas
reservoir is exhausted. There are no feedback mechanisms in our model,
as it concentrates on the interplay between gravity and gas pressure
alone.  Hence, it is scale free, and can be applied to star forming
regions of vastly different properties. When applied to a dark cloud
like Taurus-Aurigae, with densities of about $n(H_2)=100\:$cm$^{-3}$
and a temperature of $T=10\:$K, Fig.~\ref{fig:a} corresponds to a cube
of length $5.2\:$pc and a total mass of $6\,300\:$M$_{\odot}$. The
time scale is $3.3 \times 10^6\:$yrs. Comparing with a high mass star
star forming region like Orion, i.e. assuming $n(H_2)=10^5\:$cm$^{-3}$
and $T=30\:$K, these values scale to $L=0.28\:$pc and
$M=1000\:$M$_{\odot}$. The corresponding time scale is
$7.5\times10^4\:$yrs.

\begin{figure}[h]
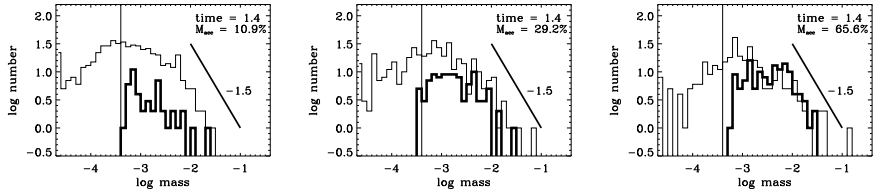

\label{fig:b}
\begin{center}
\unitlength1cm
\begin{picture}(12,2.5)
\put( 0., 0.0){\epsfxsize=4cm \epsfbox{klessen-fig-2a}}
\put( 4., 0.0){\epsfxsize=4cm \epsfbox{klessen-fig-2b}}
\put( 8., 0.0){\epsfxsize=4cm \epsfbox{klessen-fig-2c}}
\end{picture}
\caption{
Mass spectrum of cores (thick line) and of identified gas clumps (thin
lines) at times $t=0.5$, $t=0.9$, and $t=1.4$. The mass resolution for
this simulation  with $200\,000$ particles is indicated by horizontal
line at $\log_{10} M=-3.4$}
\end{center}
\end{figure}
\vspace{-0.2cm}For comparison with observational data, we plot in Fig.~\ref{fig:b}
the mass distribution of identified gas clumps in the simulation at
$t=0.5,\: 0.9$ and $1.4$ (thin lines). It agrees remarkably well with
the distribution in molecular clouds, which is analyzed as
following a power law with slope -1.5 for high clump masses (Blitz
1993); this is added as reference in the plot.  Furthermore, we show the
mass spectrum of the protostellar cores (thick lines).  

Typically, cores forming in the most massive clumps tend to populate
the high-mass end of the distribution, simply because they have the
largest gas envelope to accrete from.  However, in detail this is
influenced by an additional process. The protostellar cores form a
dense cluster, and while gaining mass via accretion, they compete with
each other for a common gas reservoir. As cluster members, they
strongly interact with each other dynamically, and as a consequence, a
considerable number of cores get expelled from their parental
envelopes into low-density regions. This terminates their growth, and
determines their position in the mass spectrum.

Under the assumption that every protostellar core forms exactly one
star of a given mass fraction, this distribution could be compared
with the stellar initial mass function. While there is reasonable
agreement for intermediate and high-mass stars, our model produces a
deficit of low-mass stars (there appears to be a plateau in the range
$-3 \le \log_{10} M \le -2$, and a steep fall-off at the resolution
limit). This is the case in all our models, regardless of the initial
conditions. This fact indicates, that for low-mass star formation
additional physical processes are important, and a detailed treatment
of protostellar accretion disks is necessary.  Furthermore, additional
fragmentation might occur in the late accretion phase leading to a
binary or multiple system within the same parental gas envelope (see
e.g. Burkert et al. 1997)

\begin{figure}
\label{fig:c}
\unitlength1cm
\begin{minipage}[b]{6cm}
\unitlength1cm
\begin{picture}(6,3.2)
\put(0,-0.9){\epsfxsize=6cm \epsfbox{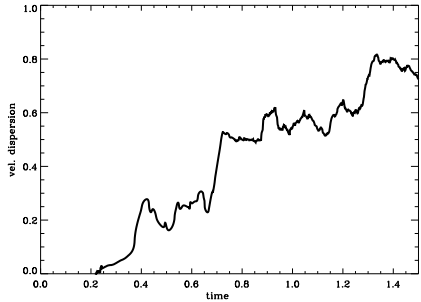}}
\end{picture}
\end{minipage}\hspace{1.0cm}
\raisebox{-0.5cm}{
\begin{minipage}[b]{5cm}
\caption{ Velocity dispersion of the newly formed cluster of
  protostellar objects as function of time. For a region like Taurus,
  the abscissae would be scaled by $1\:$km/s and the ordinates by
  $3.3\times 10^6\:$yrs. For a Orion-like region, these values are
  2.6$\:$km/s and $7.5\times 10^4\:$yrs.}
\end{minipage}}
\end{figure}

Figure~3 finally plots the velocity dispersion of the cluster.
Again, this value can be compared to the data from young stellar clusters, like in
Taurus or Orion. Despite the simplifications in our model, the kinematical
properties are in excellent agreement with the observed values.

\end{document}